\documentclass[%
 reprint,
 amsmath,amssymb,
 aps,
 prl,
]{revtex4-2}

\usepackage{graphicx}
\usepackage{dcolumn}
\usepackage{bm}

\usepackage{color, upgreek}

\definecolor{darkspringgreen}{rgb}{0.09, 0.45, 0.27}

\definecolor{deepcarrotorange}{rgb}{0.91, 0.41, 0.17}

\begin{document}

\preprint{APS/123-QED}

\title{{Boundary Curvature Effect on the Wrinkling of Thin Suspended Films}}

\author{Stoffel D. Janssens}
\affiliation{
Mathematics, Mechanics, and Materials Unit (MMMU), Okinawa Institute of Science and Technology Graduate University (OIST), 1919-1 Tancha, Onna-son, Kunigami-gun, Okinawa, Japan 904-0495
}
\author{Burhannudin Sutisna}
\affiliation{
Mathematics, Mechanics, and Materials Unit (MMMU), Okinawa Institute of Science and Technology Graduate University (OIST), 1919-1 Tancha, Onna-son, Kunigami-gun, Okinawa, Japan 904-0495
}
\author{Alessandro Giussani}
\affiliation{
Mathematics, Mechanics, and Materials Unit (MMMU), Okinawa Institute of Science and Technology Graduate University (OIST), 1919-1 Tancha, Onna-son, Kunigami-gun, Okinawa, Japan 904-0495
}
\author{David V\'azquez-Cort\'es}
\affiliation{
Mathematics, Mechanics, and Materials Unit (MMMU), Okinawa Institute of Science and Technology Graduate University (OIST), 1919-1 Tancha, Onna-son, Kunigami-gun, Okinawa, Japan 904-0495
}
\author{Eliot Fried}
\email{eliot.fried@oist.jp}
\homepage{https://groups.oist.jp/mmmu}
\affiliation{
Mathematics, Mechanics, and Materials Unit (MMMU), Okinawa Institute of Science and Technology Graduate University (OIST), 1919-1 Tancha, Onna-son, Kunigami-gun, Okinawa, Japan 904-0495
}

\date{\today}

\begin{abstract}
In this letter, we demonstrate a relation between the boundary curvature $\kappa$ and the wrinkle wavelength $\lambda$ of a thin suspended film under boundary confinement. Experiments are done with nanocrystalline diamond films of thickness $t \approx 184$~nm grown on glass substrates. By removing portions of the substrate after growth, suspended films with circular boundaries of radius $R$ ranging from approximately 30 to 811 $\upmu$m are made. Due to residual stresses, the portions of film attached to the substrate are of compressive prestrain $\epsilon_0 \approx 11 \times 10^{-4}$ and the suspended portions of film are azimuthally wrinkled at their boundary. We find that $\lambda$ monotonically decreases with $\kappa$ and present a model predicting that $\lambda \propto t^{1/2}(\epsilon_0 + \Delta R \kappa)^{-1/4}$, where $\Delta R$ denotes a penetration depth over which strain relaxes at a boundary. This relation is in agreement with our experiments and may be adapted to other systems such as plant leaves. Also, we establish a novel method for measuring residual compressive strain in thin films.
\end{abstract}

\maketitle

\begin{figure}
\centering
\includegraphics{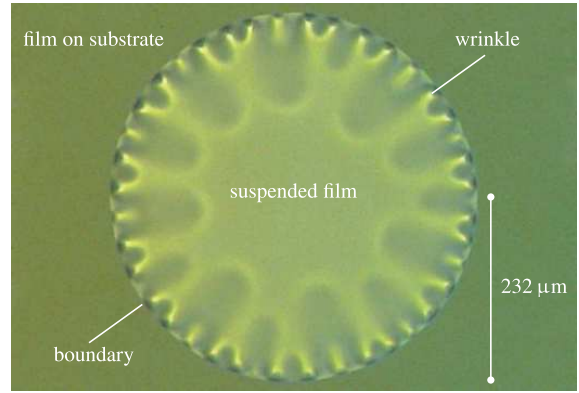}
\caption{Reflecting optical microscope image of a nanocrystalline diamond (NCD) film of approximate thickness $t = 184$~nm that was grown on a glass substrate. The suspended portion of film has a circular boundary of radius $R = 232$ $\upmu$m and was made by etching the substrate from the backside. Due to residual stresses, the portion of film attached to the substrate is of compressive prestrain $\epsilon_0 \approx 11 \times 10^{-4}$ and the suspended portion of film is azimuthally wrinkled.}
\label{fig_1}
\end{figure}

Wrinkling is an ubiquitous natural phenomenon that has led to the evolution of tissue that lowers the expenditure of energy \cite{Nixon2015,Martins2018} and is being explored by man to design efficient devices with self-similar patterns \cite{Kim2012,Li2018}. Thin films are known to wrinkle because of excess area and geometric compaction \cite{Huang2010,Leocmache2015}.
An example of this phenomenon is shown in Fig.~\ref{fig_1}, which depicts a thin suspended nanocrystalline diamond (NCD) film with circular boundary. This film was grown on a glass substrate, after which a circular portion of that substrate was removed by etching. Due to residual compressive stresses in the supported portion of the film and to the presence of boundary confinement, excess area forms and geometric compaction occurs. The strain in the film is relaxed radially by buckling and azimuthally by wrinkling. In this Letter, we present experiments showing that the boundary curvature of this type of system has a relatively strong impact on azimuthal wrinkling and provide a simple model that predicts this. We also describe a method that we developed for measuring strain in the supported film from the height profile of the suspended film. The method complements those reviewed by Abadias and co-workers \cite{Abadias2018}.
\begin{figure*}
\centering
\includegraphics{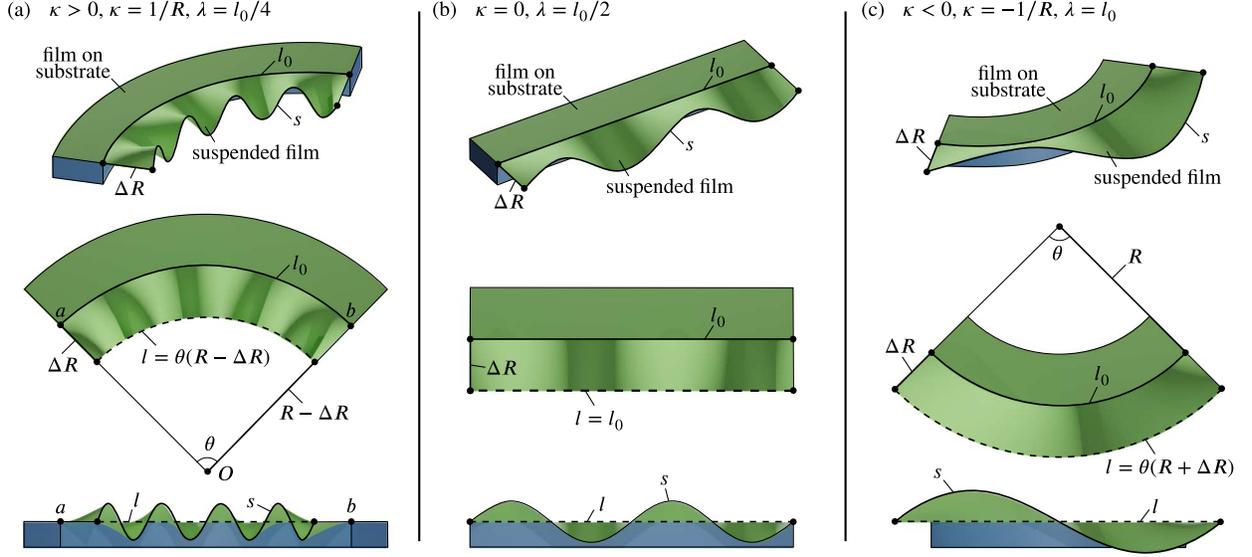}
\caption{(a) Schematic of a suspended portion of film, based on Fig.~(\ref{fig_1}), that assists in describing the model presented in this Letter. The length of a portion of boundary is denoted by $l_0$ and the penetration depth $\Delta R$ is on the order of the length over which strain in the suspended film relaxes. At distance $\Delta R$ from the boundary, a curve of arc length $l$, which is the projection of the oscillating curve of length $s$, exists. The projection is done from the suspended film, which is of radius $R$ and boundary curvature $\kappa > 0$, to the plane $\mathcal{P}$, which is defined by points $a$, $O$, and $b$. Within $\mathcal{P}$, a polar coordinate system with origin $O$, radial coordinate $r$, and azimuthal angle $\theta$ is defined. The radial distance $d$ is measured from the boundary to $O$. (b) and (c) are schematics that are similar to (a) but with $\kappa = 0$ and $\kappa < 0$, respectively. The effect of $\kappa$ on wrinkle wavelength $\lambda$ is illustrated by increasing $\lambda$ monotonically from (a) to (c).}
\label{fig_2}
\end{figure*}

The behavior of wrinkles is typically modeled using the F\"oppl--von K\'arm\'an theory of plates in conjunction with scaling arguments and asymptotic analysis. The characteristic wrinkle wavelength $\lambda$ of a thin film is found by minimizing energy and is provided by scaling relation
\begin{equation}
\lambda\sim t^{1/2}\epsilon^{-1/4},
\label{eq_lambda}
\end{equation}
where $t$ is the film thickness and $\epsilon$ is a strain that induces wrinkling \cite{Gioia1997,Cerda2003,Puntel2011}.

Efforts to investigate the effect of boundary curvature on the wrinkling of thin suspended films began only recently. Experiments and simulations have shown that curvature can significantly affect the wrinkling of cylindrical shells upon stretching \cite{Wang2020}. For spherical \cite{Li2011,Breid2013} or tubular \cite{Ciarletta2014,Yang2018} bilayer systems, similar conclusions have been reached. Still, the effect of curvature on the benchmark system that we investigate here is to the best of our knowledge not addressed in literature.

We present results from systematic experiments done with suspended NCD films of thickness $t \approx 184$~nm, nearly circular boundaries of radius $R$ ranging from 30 to 811~$\upmu$m, and a compressive prestrain $\epsilon_0 \approx 11 \times 10^{-4}$, which prevails at the boundaries. In contrast to ultra-thin polymer films \cite{Huang2007}, NCD films can be regarded as elastic. The prestrain $\epsilon_0$ is therefore time-invariant and the wrinkling and underlying strain of the films can be characterized in detail. Only through recent technological advances \cite{Janssens2019} has it become possible to systematically fabricate suspended NCD films with radii less than 250~$\upmu$m.

From the experiments described hereinafter, we find that $\lambda$ monotonically decreases with curvature $\kappa = 1/R$. To explain this trend, we present a model involving a penetration depth $\Delta R$ over which the strain in the suspended film relaxes. The setup of our problem is depicted in Fig.~\ref{fig_2}(a), where $\Delta R$ and all other salient geometrical quantities entering our formulation are described. We assume that the radial strain $\epsilon_{rr}$ is independent of the hoop strain $\epsilon_{\theta\theta}$, that the shear strain $\epsilon_{r\theta}$ vanishes, and that the film is inextensional for $d \geq \Delta R$. Since wrinkling occurs azimuthally with respect to the origin $O$, we infer that $\epsilon = \epsilon_{\theta\theta}$. On this basis, we find that
\begin{equation}
\epsilon = \frac{s}{l} - 1= \frac{s}{\theta(R - \Delta R)} - 1.
\label{eq_eps}
\end{equation}
If the curvature $\kappa$ of the supporting edge vanishes, as depicted in Fig.~\ref{fig_2}(b), we find that
\begin{equation}
\epsilon = \frac{s}{l} - 1 = \frac{s}{l_0} - 1 = \epsilon_0.
\label{eq_eps_0}
\end{equation}
For $\kappa > 0$, $l < l_0$, which shows that geometric compaction occurs when the curvature $\kappa$ of the supporting edge is positive. If $R\gg\Delta R$, we can apply the binomial theorem to expand $(R - \Delta R)^{-1}$ and write Eq.~(\ref{eq_eps}) as
\begin{equation}
\epsilon \sim \epsilon_0 + \frac{s}{l_0} \Delta R \kappa.
\label{eq_eps_app}
\end{equation}
Since $R\gg\Delta R$, the quotient $s/l_0$ is practically unity and we may write Eq.~(\ref{eq_eps_app}) as
\begin{equation}
\epsilon \sim \epsilon_0 + \Delta R \kappa.
\label{eq_eps_fin}
\end{equation}
With Eqs. (\ref{eq_lambda}) and (\ref{eq_eps_fin}), we then obtain
\begin{equation}
\lambda \sim \frac{t^{1/2}}{(\epsilon_0 + \Delta R \kappa)^{1/4}}.
\label{eq_lambda_app}
\end{equation}
For $\kappa < 0$ and $\kappa=-1/R$, as depicted in Fig.~\ref{fig_2}(c), Eq.~(\ref{eq_lambda_app}) is also obtained using similar arguments as for $\kappa > 0$. Granted that $\Delta R$ and $\epsilon_0$ are constant, Eq.~(\ref{eq_lambda_app}) predicts that $\lambda$ decreases monotonically with $\kappa$. As expected, Eq.~(\ref{eq_eps_fin}) reduces to Eq.~(\ref{eq_eps_0}) for $\kappa = 0$. 

To fabricate our films, we first seed $10 \times 10 \times 0.2$~mm$^3$ Lotus NXT glass substrates with nanodiamonds \cite{Williams2007,Janssens2011}. Subsequently, a closed film is grown with plasma assisted chemical vapor deposition in the reactor of an SDS6500X microwave system with 1.5~kW of 2.45 GHz microwaves. During film growth, the substrate temperature is maintained at about 873 K and the reactor is kept at a pressure of 2~kPa with 294~sccm of hydrogen gas and 6~sccm of methane gas. The thicknesses of the NCD films were measured with a Hamamatsu C13027 optical nano gauge. The through-glass vias and the suspended films were fabricated by etching the substrate locally with hydrofluoric acid using recently described techniques \cite{Janssens2014a,Janssens2019}. Reflecting optical microscope images and surface profiles of suspended films were taken with a Keyence VK-X150 confocal laser microscope. We define a surface profile as a collection of heights of a suspended film with reference to plane $\mathcal{P}$. Similar to previous work \cite{Janssens2019}, film stresses are obtained by X-ray diffraction measurements carried out with a Bruker D8 Discover diffractometer.

To estimate the length scale of $\Delta R$, we use the dimensionless number $\mathcal{D} = S/B$, where $S$ and $B$ denote stretching and bending stiffness, respectively. In so doing, we assume that a portion of suspended film acts as a cantilever of width $w$, thickness $t$, and length $d$. Then, $S = E t w / d$ and $B = E t^3 w / 4 d^3$, where $E$ denotes Young's modulus, so that $\mathcal{D} = 4(d/t)^2$. For $d = t$, we see that $\mathcal{D} = 4$, which indicates that bending and stretching stiffness are of the same order. However, for a relatively small value of $d$, for example 2~$\upmu$m, and $t = 184$~nm, we see that $\mathcal{D} = 473$, which indicates that bending is strongly favorable over stretching. On this basis, we estimate that $\Delta R$ is on the order of microns, that $R \gg \Delta R$, and that the suspended NCD films are practically inextensional.
\begin{figure}
\centering
\includegraphics{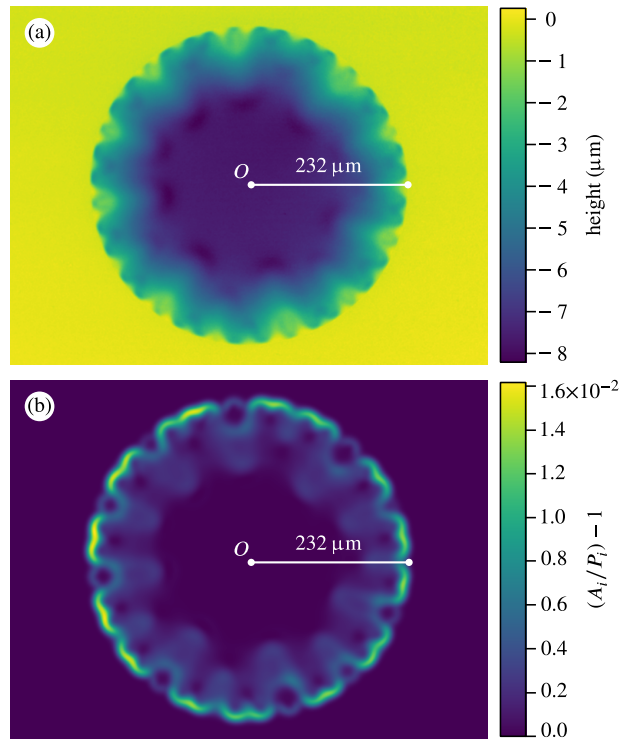}
\caption{(a) Surface profile of the suspended film depicted in Fig.~1. From the profile, it is clear that radially the film buckles out of plane $\mathcal{P}$ and azimuthally the film wrinkles. This is one of the profiles used to obtain $\epsilon_0 = (11 \pm 2) \times 10^{-4}$ with $\epsilon_0 \sim \sqrt{A/\pi R^2} - 1$. Here, $A$ denotes the surface area of the suspended film, as found by removing noise from a surface profile, creating a surface from that smooth surface profile by triangulation, calculating the area $A_i$ of each triangle $T_i$, and summing up all $A_i$ that correspond to suspended film. (b) $(A_i/P_i) - 1$ obtained from (a) plotted with respect to $\mathcal{P}$. Here, $P_i$ denotes the area of the projection of $T_i$ on $\mathcal{P}$. Granted that the suspended film is inextensional, $A_i/P_i$ is the scaled film area density. This ratio is greatest at the wrinkled portion of the suspended film.}
\label{fig_3}
\end{figure}

From XRD measurements, we infer that $\epsilon_0 = (11 \pm 2) \times 10^{-4}$; however, when the material properties of a film are unknown, or when dealing with ultra-thin films or amorphous materials, other methods are needed to obtain $\epsilon_0$. To solve this problem, we demonstrate that $\epsilon_0$ is estimated accurately by
\begin{equation}
\zeta = \sqrt{\frac{A}{\pi R^2}} -1,
\label{eq_zeta_f}
\end{equation}
where $A$ denotes the surface area of the suspended film. The relation $\epsilon_0 \sim \zeta$ holds if the suspended film is practically inextensional, a criterium that is met in our work. The area $A$ of the suspended film is estimated by removing noise from a surface  profile with a Gauss filter, creating a surface from that filtered profile by triangulation, calculating the area $A_i$ of each surface triangle $T_i$, and summing up all $A_i$ that correspond to suspended film. We find for our films that $\epsilon_0 = (11 \pm 2) \times 10^{-4}$. This value is in excellent agreement with the value obtained from XRD measurements and shows that our approach yields results consistent with a standard strain characterization method. One of the surface profiles used to obtain $\epsilon_0$ appears in Fig.~\ref{fig_3}(a), which is for the film shown in Fig.~\ref{fig_1}. Fig.~\ref{fig_3}(b) depicts $(A_i/P_i)-1$, as obtained from the height profile in Fig.~\ref{fig_3}(a), plotted with respect to $\mathcal{P}$, in which $P_i$ denotes the area of the projection of $T_i$ on $\mathcal{P}$. The ratio $A_i/P_i$ represents the scaled film area density under the assumption that the suspended film is inextensional. From the plot we observe that $A_i/P_i$ is greater at the wrinkled portion than at the remaining area of the suspended film. This shows that in addition to providing a value for $\epsilon_0$, our strain analysis is useful for investigating film area density.

\begin{figure}
\centering
\includegraphics{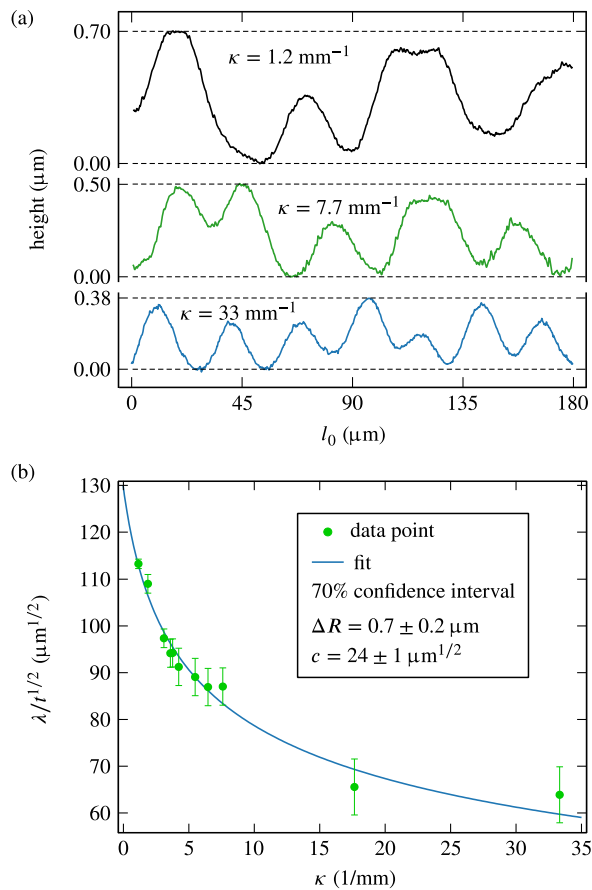}
\caption{(a) Surface profiles of suspended NCD films measured at approximately 6~$\upmu$m from their boundaries, with respect to $\mathcal{P}$, and plotted versus $l_0$. A relatively small offset in height is applied for clarity. From this data, we deduce that the wrinkle wavelength $\lambda$ and amplitude $A$ decrease monotonically with $\kappa$. (b) Wrinkle wavelength $\lambda$ divided by $t^{1/2}$ versus $\kappa$. Eq.~\ref{eq_lambda} is fitted to the data with $\epsilon_0 = 11 \times 10^{-4}$\ and fitting parameters $c$ and $\Delta R$, with $c$ denoting a proportionality constant. This result supports the assertion that our wrinkling model captures the main physical ingredients that explain our observations.}
\label{fig_4}
\end{figure}

To demonstrate that $\lambda$ decreases monotonically with $\kappa$, three surface profiles of suspended NCD films taken at $d \approx 6$ $\upmu$m are given in Fig.~\ref{fig_4}(a), with a relatively small offset in height for clarity. Accurate values of $\lambda$ are found by counting the number of wrinkles of a suspended film and dividing the resulting number by $2\pi R$. Counting is most easily done with reflecting optical microscope images, of which one appears in Fig.~\ref{fig_1}. For that image, $39 \pm 1$ wrinkles were counted, as confirmed from analyzing height profiles such as those given in Fig.~\ref{fig_4}(a). In Fig.~\ref{fig_4}(b), the obtained values of $\lambda$, scaled by $t^{1/2}$, are plotted versus $\kappa$. For $\epsilon_0 = 11 \times 10^{-4}$, $\Delta R$ assumed to be constant, and $\Delta R$ and proportionality constant $c$ acting as fitting parameters, we fit Eq.~(\ref{eq_lambda_app}) to the data in Fig.~\ref{fig_4}(b) with the least-squares method. The fit confirms that our wrinkling model captures the main physical ingredients that explain our observations. For a confidence interval of 70\%, we find that $c = 24 \pm 1$ $\upmu$m$^{1/2}$ and $\Delta R = 0.7 \pm 0.2$ $\upmu$m. This results supports the relation $R \gg \Delta R$.

Within the framework of our model, the arclength of the curves in Fig.~\ref{fig_4}(a) should be similar as a consequence of inextensionality, in which case the wrinkle amplitude decreases with the wrinkle wavelength $\lambda$. This is verified by comparing the height axis of the surface profiles in Fig.~\ref{fig_4}(a). Further analysis of $A$ versus $d$ may be done within the formalism based on wrinklons \cite{Vandeparre2011}, which we leave for future work.

For a leaf that is supported by a stem of radius $R$, Xu and coworkers \cite{Xu2020} suggest that residual growth strain induces wrinkling of the leaf. Interestingly, they predict that the associated wrinkle wavelength $\lambda$ decreases monotonically with $R$. By modeling such a leaf as a thin suspended film supported by a rigid stem of curvature $\kappa = -1/R$, as depicted in Fig.~\ref{fig_2}(c), and attributing $\epsilon_0$ to growth, our simple model provides a similar prediction, granted that $\Delta R$ and $\epsilon_0$ are approximately constant. Experimentally, the case $\kappa < 0$ may be verified by fabricating micro-disk like structures \cite{Sartori2018}. 

To finalize our discussion, we underline that due to the presence of curvature in our experiments the wrinkle density is practically doubled. It is therefore evident that, apart from dynamic wrinkling {\cite{Box2019}, boundary curvature also needs to be taken into account when designing devices with functional wrinkles.

We conclude that boundary curvature can strongly influence the wrinkling of suspended films. Experimentally we showed this by growing nanocrystalline diamond films of approximate thickness 184~nm on glass substrates. Due to residual stresses, a compressive strain in the films is introduced. By removing portions of the substrate, suspended, azimuthally wrinkled films with circular boundaries of approximate radius 30--811~$\upmu$m were made. We found that the wavelength of these wrinkles decreases monotonically with boundary curvature, leading to a doubling of the wrinkle density. To explain this, we provided a simple model that is in line with our experiments and may be adapted to other systems such as suspended plant leaves. Additionally, taking advantage of the fact that thin suspended films can be regarded as inextensional, we established a novel method for measuring residual compressive strain and film area density from height profiles.
\begin{acknowledgments}
We gratefully acknowledge the support from the Okinawa Institute of Science and Technology Graduate University with subsidy funding from the Cabinet Office, Government of Japan.
\end{acknowledgments}



%
\end{document}